\documentclass[journal,onecolumn]{IEEEtran}

\input epsf
\usepackage{cite}
\usepackage{hyperref}
\usepackage{xcolor}

\usepackage[pdftex]{graphicx}

\usepackage{url}       
\begin{document}

\title{\LARGE Thermodynamic Cost of Edge Detection in Artificial Neural Network (ANN)-Based Processors}

\author{Se\c{c}kin Bar{\i}\c{s}{\i}k and  \.{I}lke Ercan\\

\thanks{During this research, both authors were affiliated with the Electrical \& Electronics Engineering Department
Bo\u{g}azi\c{c}i University
\.{I}stanbul, Turkey. 
Dr. \.{I}lke Ercan is currently affiliated with the Engineering Department at University College Roosevelt in Middelburg, the Netherlands. Corresponding author of the paper is Se\c{c}kin Bar{\i}\c{s}{\i}k. E-mail: seckin.barisik@boun.edu.tr. }
\thanks{This research is supported in part by Bo\u{g}azi\c{c}i University BAP Start-up Grant No: 11540.}

}



%


\maketitle

\begin{abstract}
Architecture-based heat dissipation analyses allows us to reveal fundamental sources of inefficiency in a given processor and thereby provide us with roadmaps to design less dissipative computing schemes independent of technology-bases used to implement the processor. 
In this work, we study architectural-level contributions to energy dissipation in Artificial Neural Network (ANN)-based processors that are trained to perform edge detection task. We compare the training and information processing cost of ANNs to that of conventional architectures and algorithms using 64-pixel binary image. Our results reveal the inherent efficiency advantages of ANN networks trained for specific tasks over general purpose processors based on von Neumann architecture. We also compare the proposed performance improvements to that of CAPs and show the reduction in dissipation for special purpose processors. 
Lastly, we calculate the change in  dissipation as a result of change in input data structure and show the effect of randomness on energetic cost of information processing. The results we obtain provide a basis for comparison for task-based fundamental energy efficiency analyses for a range of processors and therefore contribute to the study of architecture-level descriptions of processors and thermodynamic cost calculations based on physics of computation.
\end{abstract}

%
\IEEEpeerreviewmaketitle

\section{Introduction}

  Overcoming energy efficiency limitations plays a key role in maintaining increased performance trends in emerging computers. 
Architecture-based heat dissipation analyses allow us to reveal fundamental sources of inefficiency in a given processor and thereby provide us with roadmaps to design less dissipative computing schemes independent of technology-bases used to implement the processor \cite{anderson,jared}. The limitations imposed on the energetic cost of information processing using conventional processor structures requires us to study energy dissipation aspect of the recent processor proposals.\par

Processors based on ANN structures propose practical solutions to various limitations involved in conventional processors including parallel processing and noise tolerance. However, ANNs are mostly studied form the perspective of parameter optimization and structure selection. In order design networks that help minimize energy dissipation and choose appropriate data sets for these networks we also need to study the thermodynamic aspect of ANNs. A recent study presented in Ref. \cite{natesh} addresses the thermodynamic aspect of feedforward neural networks using a single layer perceptron model. This study, focusing on minimizing the dissipative cost of training, presents the fundamental lower bounds on energy dissipation for a single perceptron designed to perform AND gate function as a illustrative example. In this work, we perform a thermodynamic analysis on a Multilayer Perceptron (MLP) by calculating the  lower bounds on edge detection task. The results we obtained provide insights into the ultimate performance limits of neuromorphic systems and understand sources of inefﬁciency included in the structure of processors and data sets.

In this paper, we calculate fundamental lower bounds on architecture-induced energy dissipation in the Artificial Neural Network (ANN) processors and compare results with von Neumann and Cellular Array Processors (CAPs). Results reveal the inherent efficiency advantages of ANN networks trained for specific task over processors based on von Neumann architecture, and we also compare the proposed performance improvements to  that  of  CAPs. The fundamental sources of inefficiency of two-layered feed-forward neural network, which is independent of technology is analyzed for edge detection task. This paper is organized as follows: We begin by introducing the structure of the ANN we employed as an illustrative example, as well as the data set generation and training methods. We then study the energetic cost of information processing in the presented network based on fundamental lower bounds, and present the energetic cost of each epoch for the trained ANN as well as processing the network. We also study the effect of input data structure on the fundamental lower bounds and calculate how randomness effects the energetic cost for information processing.  We conclude by final remarks and future directions.

\section{ANNs and Edge-Detection}

ANNs are inspired by biological systems that are capable of learning and recognizing patterns of data as a part of their neural processing. Unlike conventional general purpose von Neumann processors (vNp) that perform computation sequentially, ANNs can perform many tasks with significant speed advantage due to their parallel processing nature.\cite{haykin} Despite many applications of ANNs, including reprensenting Boolean functions \cite{hertz}, translating written English text into spoken text \cite{james} edge-detection task has not been proven to be ideal operation to illustrate ANNs capability, however, literature contains edge-detection-based comparison of vNps and CAPs, and in order to provide a comparative analysis, we study energy efficiency of ANNs in edge-detection. ANNs applied to edge detection task provide solution to problems where algorithmic methods are computationally intensive and noise-intolerant\cite{becerikli}. In this work, we employ a simple and widely used neural network structure focusing on the Multilayer Perceptron (MLP) network using backpropagation learning technique.\par 

Feedforward neural networks, including MLPs, contain an input layer, one or more hidden layers, and an output layer all connected with synaptic weights. This allows these networks to overcome the practical limitations of single layer perceptrons including nonlinear function approximation. In MLPs, training has two phases; forward and backward propagation. In the forward propagation, synaptic weights are fixed and information flows from the input to the output. It is used to get output and to compare it with real (or target) value to get an error. In the backpropagation, the error flows backward direction in this time and error is tried to minimize by adjusting synaptic weights.\par

In the following section, we introduce the structure of the ANN used to perform edge detection task in this paper, and discuss the mathematical function performed by the network. The training method used as well as the performance of the output is discussed as an illustrative example.

\section{Illustrative Example}
We employ a feedforward ANN with back-propagation \cite{rumelhart} performing edge-detection for nine-pixel patches prepared from entire image. Here, we present the structure of the network and the mathematical representation of the task performed as well as the training method and the output performance.
\subsection{Structure of the ANN}
 The structure of the ANN architecture used as an illustrative example in our study is depicted in Fig. 1. The input layer consists of  9 neurons, the hidden layer includes 12 neurons, and the output consists of a single neuron. Each neuron in the network has a specific synaptic weight associated with the image that will be used for edge-detection. We employ the ANN visualizer provided in Ref. \cite{visualizer} to plot the network in Fig. 1.  The mathematical representation of the tasks performed by this network is provided by the activation functions.

\begin{figure}
    \centering
    \includegraphics[width=0.5\textwidth]{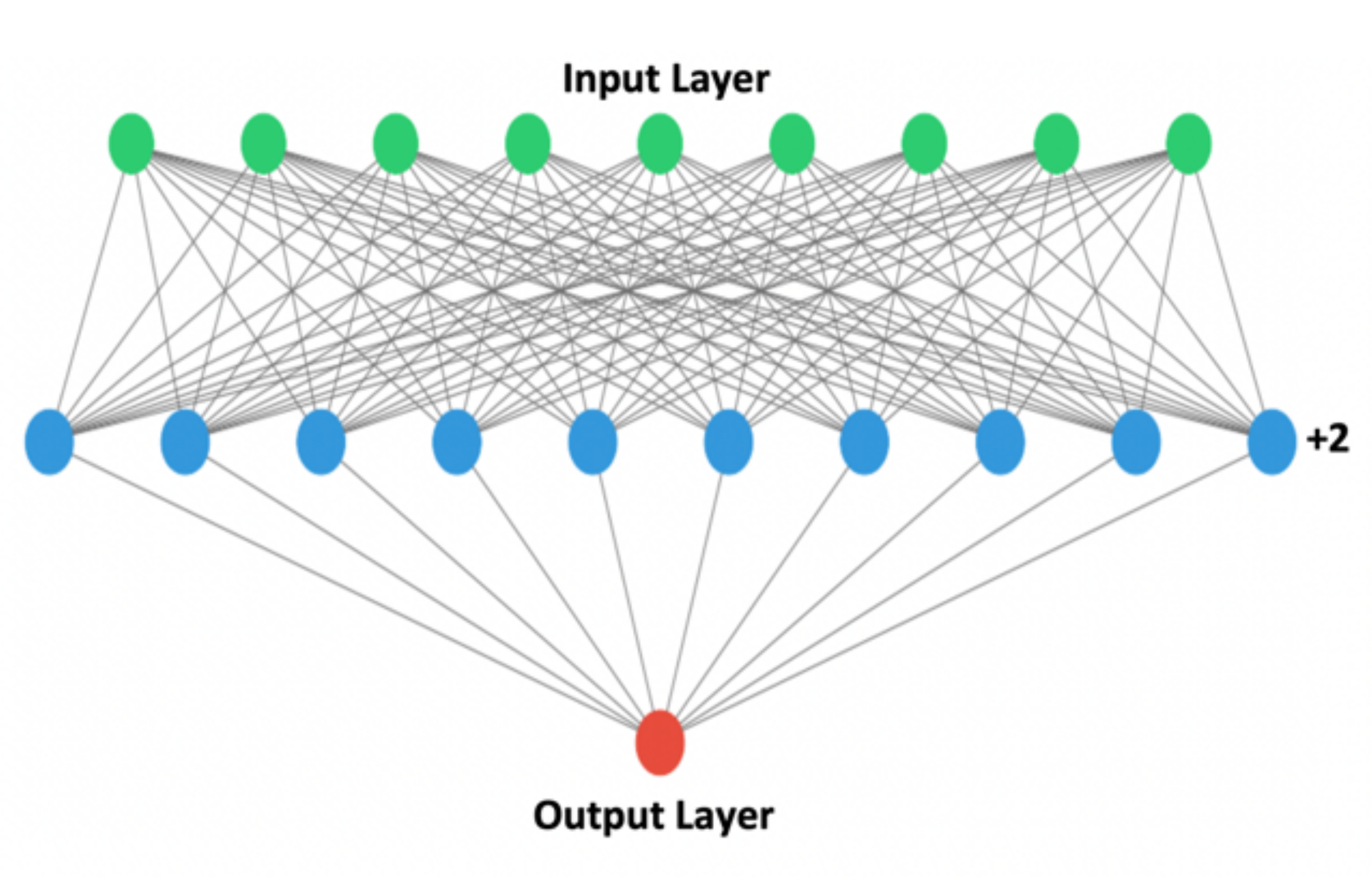}
    \caption{ANN architecture  performing edge-detection task developed using ANN visualizer. }
\end{figure}

\subsection{Activation Function}
The network presented in Fig. 1 propagates information based on the activation functions shown in Fig. 2.  The mathematical function chosen to propagate the input to the hidden layer is a Rectiﬁed Linear Unit (ReLU) activation function, 
\begin{equation}
 f_R(z) = max (0, z). 
\end{equation}
Once the information moves from the input layer to hidden layer, we then use a sigmoid activation function to obtain the output,
\begin{equation}
  f_S(z) = \frac{1}{ 1 + e^{-z} }.
\end{equation}
Eq. 2 is a suitable choice for binary classification task as the sigmoid function is between 0 and 1.  ReLU, used in the first step, allows for faster training of neural networks on large and complex data sets, and it has fewer vanishing gradient problems compared to sigmoid function. However, any negative input is turned to zero by the ReLU activation function means inappropriate mapping of the negative values and it causes inefficiency. 
\begin{figure}[h!]
    \centering
    \includegraphics[width=0.5\textwidth]{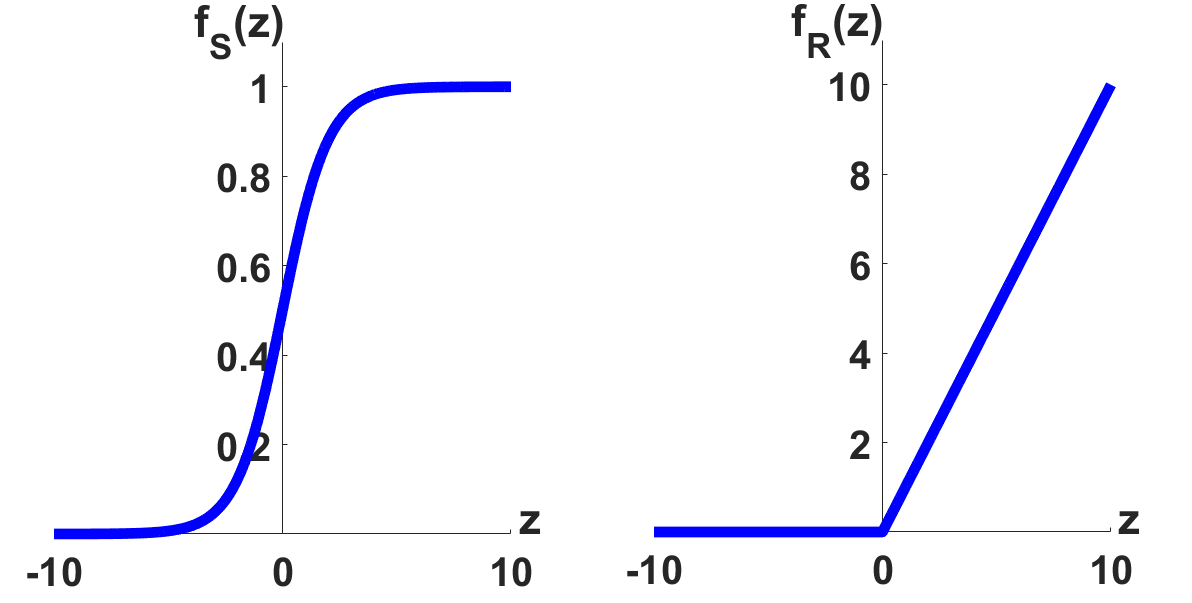}
    \caption{Sigmoid and ReLU activation functions}
    \label{architecture}
\end{figure}
\begin{figure}[h!]
    \centering
    \includegraphics[width=0.5\textwidth]{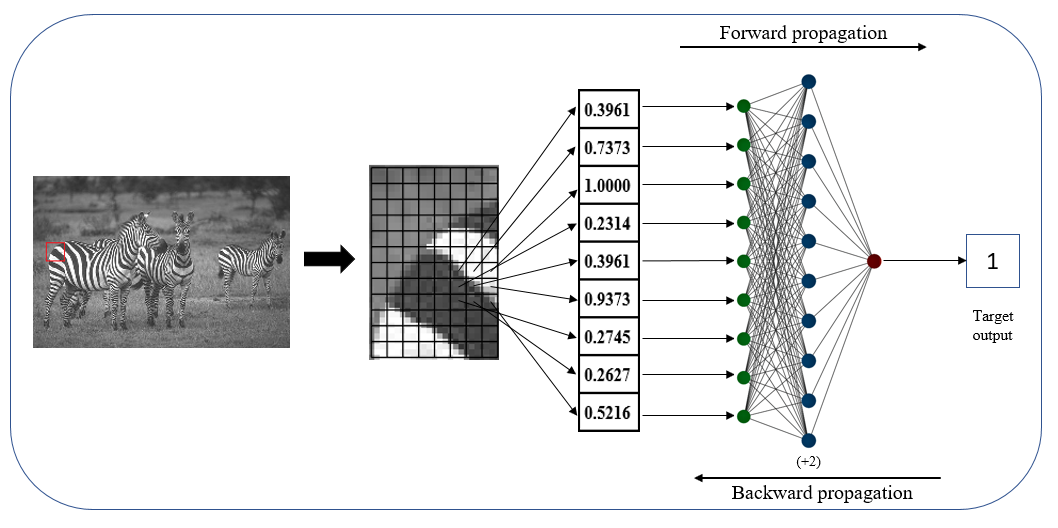}
    \caption{Training method of ANN}
\end{figure}

\subsection{Training Method}
In order to perform edge-detection task, we train the network using 20 images (16 for training set and 4 images for validation set) from the BSDS300 dataset \cite{Berkeley} over 100 epochs as shown in Fig. 3. The input images from the BSDS300 set are gray-scale pixels whose values ranging from 0 to 255 and whose sizes are 481x321. Data is prepared as
3x3 image patches and corresponding label which is center of 3x3 window of edge map human annotations \cite{wang}. Weights are initialized with He initialization \cite{He}, mini-batch size is selected 128, mean squared error (MSE) is used as loss function. All training is done by use of back-propagation learning rule with Adam optimization algorithm \cite{kingma} which speed up convergence. Early stopping method, stopped the training at the point where validation error started increasing to avoid over-fitting, is used. Training and validation error curves during the training are provided in Fig. 4. \cite{early}
\begin{figure}[h!]
    \centering
    \includegraphics[width=0.5\textwidth]{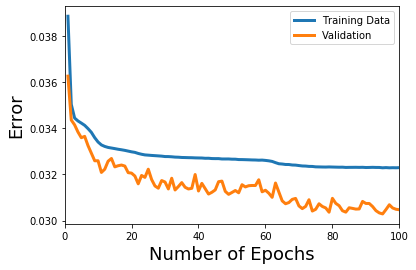}
    \caption{Training and validation set errors of the ANN}
\end{figure}

\par We tested our ANN edge detector using the some images, which weren't seen by the ANN, in BSDS300 dataset. Results of the test are shown in Fig. 5. Here, our focus is not optimization but rather analysis of state evaluations. In literature, with different epoch number and different data set, common error values are about 1\% \cite{becerikli}.

\begin{figure}[h!]
    \centering
    \includegraphics[width=0.5\textwidth]{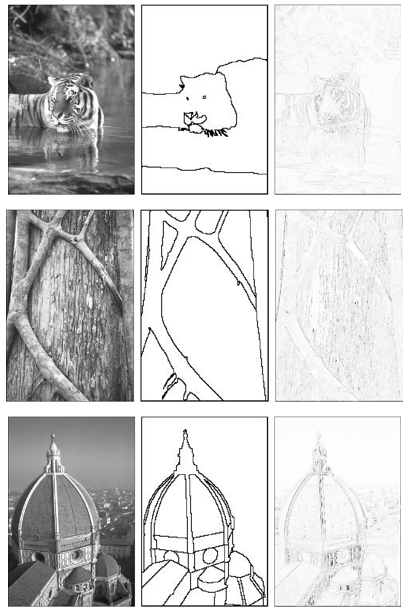}
    \caption{Results of the tests. The first and second column represents the original image and the ground truth, respectively. The third column shows the raw output of our ANN.}
\end{figure}

\begin{figure}[h!]
    \centering
    \includegraphics[width=0.5\textwidth]{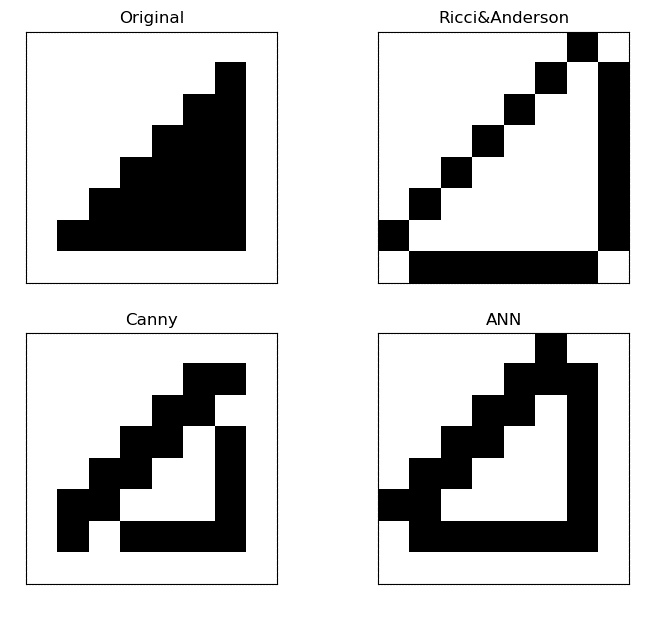}
    \caption{The original image and the edge-detection output using three different approaches.}
    \label{comparison}
\end{figure}

\subsection{Comparison of the Output}
In order to provide a perspective on the performance of edge-detection task based on different approaches, we have chosen the image employed in Ref. \cite{jared}. The original image, as well as three schemes used to perform the detection are shown in Fig. 6. The output obtained in Ref. \cite{jared} is shown on top right. We also performed edge detection on the same image using conventional Canny algorithm as shown in bottom left. Lastly, the result of the edge detection using our ANN is shown in bottom right. The three outputs shown in the figure reflect the change in accuracy and precision of the edge-detection task. 
Here we only consider first 10 epochs since the image becomes distinct early on in the training and the information-theoretic content of the states are tractable. 

\section{Fundamental Lower Bounds}

The fundamental lower bounds obtained via using processor thermodynamic methodology similar to those presented in Ref. [1], [2] and [3] give us technology independent lower limits on dissipation resulting from the unavoidable irreversible information loss. Such a bound can be represented as  [1]
\begin{equation}
    \Delta E \geq k_BT\ln(2) H(X|Y)
\end{equation}
where k$_B$ is Boltzmann constant, $T$ is temperature and H(X$\mid$Y) is the conditional entropy reflecting the irreversible loss that takes place during information processing. In our illustrative example, we take this entropy to be the conditional entropy between the two steps of information flowing  (1) from input to hidden layer and (2) from hidden layer to output. \par 

We obtain the fundamental lower bounds by calculating the amount of information that is irreversibly lost at each epoch of ANN operation. The cumulative lower bound energy for the ANN that is trained specifically for the edge detection task is:
\begin{equation}
    \Delta E \geq 2.0574k_BT\ln(2)
\end{equation}

It is important to note here that this bound reflects purely the energy that stems from loss of correlation between steps of edge-detection task implemented using ANN. Any dissipation resulting from the underlying physical operations to perform computation is not included in this fundamental lower bound.

\begin{figure}[h!]
    \centering
    \includegraphics[height=0.26\textwidth]{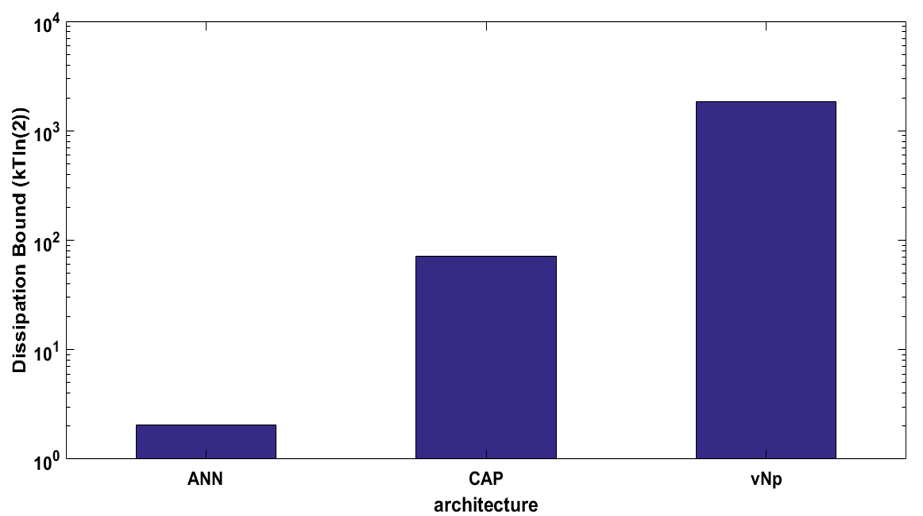}
    \caption{Dissipation bound comparison of three different processor architecture.}
    \label{arch_comparison}
\end{figure}

Our calculations show that the majority of information is lost during the calculation of the ReLU of its input.
In Fig. 7, we compare this bound with the results presented in Ref. [2] where they obtain a $1856k_B T \ln(2)$ for vNp and $71k_B T\ln(2)$ for CAP; i.e. the dissipation bound of general purpose von Neuman processor is 902 times and special purpose CAP is $34.5x$ times larger than the bound obtained for the trained artificial neural network. 

\begin{figure}[h!]
    \centering
    \includegraphics[width=0.51\textwidth]{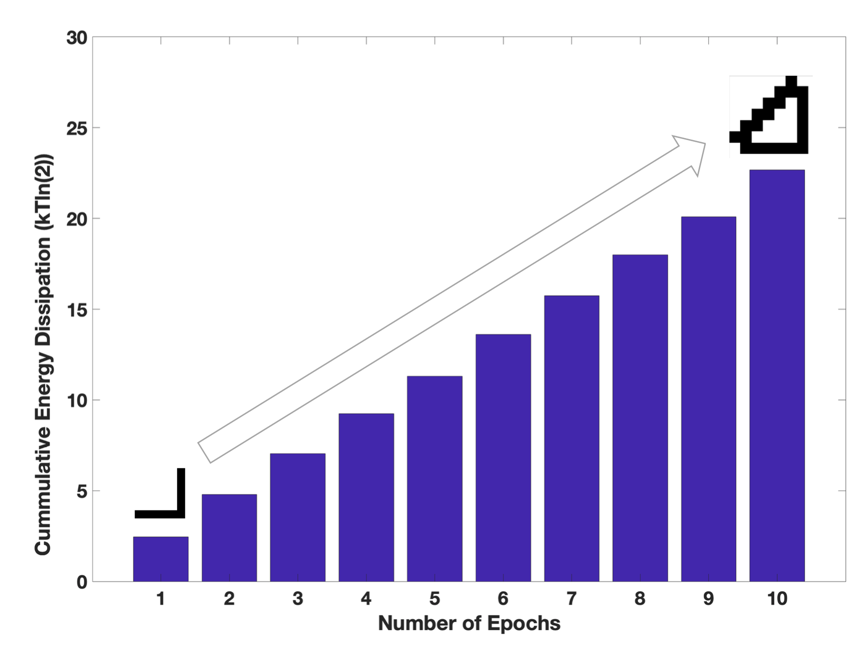}
    \caption{Cumulative Energy Dissipation based on number of epochs.}
    \label{cumulative}
\end{figure}
This comparison reflects the energetic cost of edge-detection task alone. 
However, in order to perform this detection, we also need to train the neural network; i.e. in order to obtain the result presented in Ref. [2] the neural network must be trained with more images and larger number of epochs. Each epoch represents one iteration forward and back propagation over the entire training set for the ANN network, and as the number of epochs increases, the output error of ANN generally decreases and performance of neural network improves, approaching the results in Ref. [2]. In Fig. 8 we provide a cumulative energy dissipation for increasing number of epochs. Performance of neural network improves at the end of each epoch in exchange for increasing cost.
Fig. 8 illustrates that the energy dissipated during the training of the network to perform the task illustrated, is roughly $10x$ times more than performing the task itself.  

\begin{figure*}[tb]
\centering
    \includegraphics[width=1.05\textwidth]{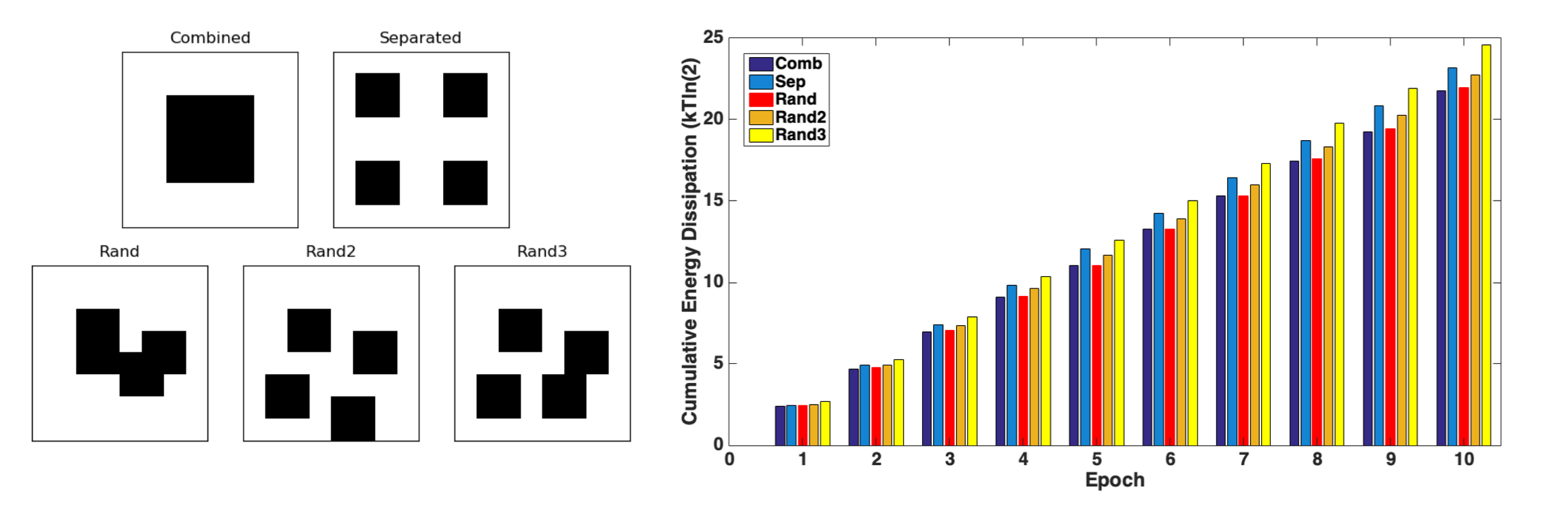}
\caption{Cumulative energy dissipation based on number of epochs for input image with varying randomness and size.}
\end{figure*}

\section{The Effect of Input Data Structure}
Our results illustrate how special purposeness of ANNs provide a significant advantage in reducing the amount of dissipation for a given task; from fundamental efficiency point of view, it requires more energy to be a general purpose processor. This shows us the importance of architecture structure in approaching fundamental lower limits. However, we also argue that, in addition to the structure of architecture, the structure of the input data set may also play a role in the amount of unavoidable energy dissipation as a result of loss of correlation during information erasure. Therefore, in addition to the effect of the processor structure, we also focus on the effect of the data set structure on the thermodynamic cost of edge-detection. \par 
In order to explore the effect of the data structure on energy dissipation, we prepared five different black and white images as shown in Fig. 9 (left) each with varying degrees of randomness involving four black squares presented in combined, separated and three randomly distributed patterns. We performed edge-detection on each of these images and present the change in energy dissipation on the cost of training based on the data structure as illustrated in Fig. 9 (right).
Cumulative energy dissipation analysis of the ANN training suggests that if the black squares are merged into a single shape this leads to a decrease in the energy dissipation. Also, for separated individual black squares, the further the images are apart from one another the higher the energy required to train the ANN for edge-detection task. We also observe that the uniform distribution of the separate dark regions does not affect the energy dissipation as much as the distance between the regions. If the shape of the separate dark regions is not identical, the energy dissipation also increases. This suggest that the randomness and the size of the image both have an effect in increasing the energy dissipated to detect the edge of the image.


\section{Conclusion}
In this work, we study the thermodynamic cost of edge-detection in ANN-based processors using MLP. We provide a comparative analysis on the fundamental energy bounds of edge-detection task as well as the cost associated with training the network and simulate the results using Python and MATLAB. The comparison we provide using three processors based on different architectures show that trained ANN have orders of magnitude lower energy requirement for edge-detection task.  Our results  support the  trade-off between energy efficiency and “general purposeness” in processors which is deeply rooted in  physics of computation. This simply suggests that even if processors were to be operated under ideal conditions, there will still be an underlining inefficiency based on the dynamics of information processing in the computing approach. Here, we show that special purposeness in ANN-based structures can help minimize this cost. \par 

In addition to the structure of the network, we also study the effect of data structure in the amount of energy dissipated to detect the edges. We show that the randomness of the data set as well as the size of the image plays a role in the amount of energy dissipated to train the ANNs.\par 

It is important to add that the fundamental lower bounds we obtain here are independent of technology-based used to implement these processor, and is based on architecture thermodynamics alone. The analyses we present here provide us insights on architectural efficiency that has physical roots and to comprehend energy costs. This comprehension can help us design computing schemes that can approach ultimate performance limits for various applications obtained using a physical-information-theoretic methodology shown here.


\section*{Acknowledgment}
This research is supported in part by Bo\u{g}azi\c{c}i University BAP Start-up Grant No: 11540. The authors would like to thank Dr. Natesh Ganesh for insightful comments on the manuscript. 




%

\end{document}